\documentclass[english,aps,prl,preprintnumbers,amsmath,amssymb,twocolumn,superscriptaddress,showpacs,floatfix]{revtex4}
\usepackage[T1]{fontenc}
\usepackage[latin9]{inputenc}
\usepackage{mathrsfs}
\usepackage{amsmath}
\usepackage{amssymb}
\usepackage{graphicx}
\usepackage{esint}

\makeatletter
\@ifundefined{textcolor}{}
{%
 \definecolor{BLACK}{gray}{0}
 \definecolor{WHITE}{gray}{1}
 \definecolor{RED}{rgb}{1,0,0}
 \definecolor{GREEN}{rgb}{0,1,0}
 \definecolor{BLUE}{rgb}{0,0,1}
 \definecolor{CYAN}{cmyk}{1,0,0,0}
 \definecolor{MAGENTA}{cmyk}{0,1,0,0}
 \definecolor{YELLOW}{cmyk}{0,0,1,0}
}

\makeatother

\usepackage{babel}
\begin{document}

\title{Vortical fluid and $\Lambda$ spin correlations in high-energy heavy-ion collisions}

\author{Long-gang Pang}

\affiliation{Frankfurt Institute for Advanced Studies, Ruth-Moufang-Strasse 1,
60438 Frankfurt am Main, Germany }

\author{Hannah Petersen}

\affiliation{Frankfurt Institute for Advanced Studies, Ruth-Moufang-Strasse 1,
60438 Frankfurt am Main, Germany }

\affiliation{Institute for Theoretical Physics, Goethe University, Max-von-Laue-Strasse 1, 
60438 Frankfurt am Main, Germany}

\affiliation{GSI Helmholtzzentrum f\"ur Schwerionenforschung, Planckstr. 1, 64291 Darmstadt, Germany}

\author{Qun Wang}

\affiliation{Interdisciplinary Center for Theoretical Study and Department of
Modern Physics, University of Science and Technology of China, Hefei,
Anhui 230026, China}

\author{Xin-Nian Wang}

\affiliation{Key Laboratory of Quark and Lepton Physics (MOE) and Institute of
Particle Physics, Central China Normal University, Wuhan, 430079,
China}

\affiliation{Nuclear Science Division, MS 70R0319, Lawrence Berkeley National
Laboratory, Berkeley, California 94720}

\begin{abstract}
Fermions become polarized in a vortical fluid due to spin-vorticity coupling. The spin polarization density is proportional to the local fluid vorticity at the next-to-leading order of a gradient expansion in a quantum kinetic theory.
Spin correlations of two $\Lambda$-hyperons can therefore reveal the vortical structure of the dense matter in high-energy heavy-ion collisions. We employ a (3+1)D viscous hydrodynamic model with event-by-event fluctuating initial conditions from A MultiPhase Transport (AMPT) model to calculate the vorticity distributions and $\Lambda$ spin correlations.
The azimuthal correlation of the transverse spin is shown to have a cosine form plus an offset due to a circular structure of the transverse vorticity around the beam direction and global spin polarization. The longitudinal spin correlation shows a structure of vortex-pairing in the transverse plane due to the convective flow of hot spots in the radial direction. The dependence on colliding energy, rapidity, centrality and sensitivity to the shear viscosity are also investigated.

\end{abstract}


\maketitle

{\it Introduction.} -- Low-energy nuclear reactions can create rotating and deformed compound nuclei that carry a large amount of orbital angular momentum of the colliding nuclei \cite{Janssens:1991ug}.  The large orbital angular momentum in non-central high-energy heavy-ion collisions cannot produce a rotating quark-gluon plasma because of the soft equation of state (EoS). It should instead lead to fluid shear and non-vanishing local fluid vorticity  \cite{Liang:2004ph,Liang:2004xn,Gao:2007bc,Becattini:2007zn,Betz:2007kg,Becattini:2007sr,Huang:2011ru,Csernai:2013bqa,Becattini:2013vja,Becattini:2015ska,Xie:2015xpa,Jiang:2016woz,Deng:2016gyh}. In such a vorticular fluid, the spin-orbital coupling polarizes the spin of fermions (quarks and baryons) \cite{Liang:2004ph,Liang:2004xn,Gao:2007bc,Becattini:2007zn,Betz:2007kg,Becattini:2007sr,Huang:2011ru,Csernai:2013bqa,Becattini:2013vja,Becattini:2015ska,Xie:2015xpa} along the direction of the vorticity. 

The mechanism of fermion polarization in a vorticular fluid is very similar to the Chiral Vortical Effect \cite{Kharzeev:2007jp,Fukushima:2008xe,Son:2009tf,Kharzeev:2010gr,Pu:2010as,Gao:2012ix}. The axial current induced by vorticity leads to the Local Polarization Effect \cite{Gao:2012ix} as a result of the spin-vorticity coupling for chiral or massless fermions \cite{Gao:2015zka}. Considering the next-to-leading order in the gradient expansion for massive fermions, one can show that the fermion spin polarization density is directly proportional to the local vorticity \cite{Becattini:2013fla,Fang:2016vpj}.  Consequently,  final state baryons such as hyperons should also be polarized along the direction of the local fluid vorticity at the freeze-out hyper-surface. Measurements of the final-state hyperon polarization, which are feasible through the parity-violating decay \cite{Abelev:2007zk}, will shed light on the vortical structure and transport properties of the strongly coupled quark-gluon plasma (sQGP) in high-energy heavy-ion collisions. 

In this Letter, we investigate the vortical structure of the sQGP in high-energy heavy-ion collisions in a (3+1)D viscous hydrodynamic model with event-by-event fluctuating initial conditions from the AMPT model \cite{Lin:2004en}. We show that the transverse vorticity of the sQGP has a circular structure around the beam direction due to the convective longitudinal flow in addition to the global alignment along the direction of the orbital angular momentum of non-central collisions. The longitudinal vorticity, however, has a vortex-pairing structure in a given transverse plane due to the convective radial flow of hot spots. We propose to use the spin correlation of two hyperons to study these vortical structures of dense matter. We will calculate hyperon spin correlations in the azimuthal angle and study their dependence on collision energy, rapidity, centrality and the shear viscosity. We neglect the spin polarization due to magnetic fields in this study.

{\it Fermion polarization in a vortical fluid.} -- Interactions in a medium with local vorticity polarize a fermion's spin due to spin-orbital coupling \cite{Liang:2004ph}.   In thermal equilibrium such a coupling between spin and local vorticity effectively shifts the energy level of fermions with different spin states. This will lead to different phase space distributions for fermions with different spin states and therefore spin polarization along the direction of the local vorticity \cite{Becattini:2013fla}. One can calculate the spin polarization in thermal equilibrium within a quantum kinetic approach \cite{Fang:2016vpj}. 

In the quantum kinetic theory, the fermion distribution is described by the Wigner function $W(x,p)$ in space-time $x$ and
4-momentum $p$,
\begin{equation}
\label{wigner}
W_{\alpha\beta}(x,p) = \int\frac{d^4 y}{(2\pi)^4}
e^{-ip\cdot y} \langle \bar\psi_\beta(x+\frac{1}{2}y) \psi_\alpha(x-\frac{1}{2}y) \rangle,
\end{equation}
where $\psi (x)$ and $\bar{\psi} (x)$ are fermionic fields, $\langle \hat{O} \rangle$ denotes the ensemble average of the operator $\hat{O}$. Note that in Eq. (1) we have neglected the electromagnetic interaction. 
Using the Dirac equation for a fermion with mass $m$, one can derive the quantum kinetic equation for the fermion's Wigner function
\cite{Elze:1986qd,Vasak:1987um},
\begin{equation}
\label{eq-c}
\left[ \gamma_\mu\left( p^\mu +\frac{i}{2} \hbar \partial^\mu_x \right) -m \right] W(x,p)=0 .
\end{equation}

The Wigner function is a $4\times4$ matrix in Dirac space and can be decomposed into 16 independent generators of the Clifford algebra. The coefficients correspond to the scalar, pseudoscalar, vector, axial vector and tensor components, respectively. The kinetic equation for the Wigner function in Eq.~(\ref{eq-c}) will lead to a system of equations for these components which can be solved through a gradient expansion. The spin polarization is given by the axial component of the Wigner function. At the next-to-leading order in gradient expansion, one obtains \cite{Fang:2016vpj} the polarization density for on-shell fermions
\begin{eqnarray}
\Pi^{\mu}(x) & = & \hbar\frac{\omega^{\mu}}{2}
\beta\int \frac{d^3p}{(2\pi)^{3}}f_{\rm FD}(x,p)\left[ 1-f_{\rm FD}(x,p)\right] ,\label{eq:spin-mass}
\end{eqnarray}
where the vorticity is defined as $\omega^{\mu}=\tilde{\Omega}^{\mu\nu}u_\nu/\beta$, 
$\tilde{\Omega}^{\mu\nu}=\frac{1}{2}\epsilon^{\mu\nu\rho\sigma}\partial_{\rho}(\beta u_{\sigma})$,
$f_{\rm FD}(x,p)$ is the Fermi-Dirac distribution,
\begin{equation}
f_{\mathrm{FD}}(x,p)=\frac{1}{e^{\beta[u\cdot p \mp\mu]}+1},
\end{equation}
for fermions ($-$) and anti-fermions ($+$) on their mass-shell ($p^2=m^2$) with the chemical potential $\mu$,  the temperature $T=1/\beta$ and
the fluid 4-velocity $u_\mu$. The energy splitting between two spin states due to spin-vorticity coupling is proportional to local vorticity. Therefore, the spin polarization density is proportional to the vorticity vector and the fermion number susceptibility at the next-to-leading-order in gradient expansion. For a finite chemical potential, the polarization per particle for fermions is always smaller than for anti-fermions \cite{Fang:2016vpj}.

Assuming the hydrodynamic evolution of dense matter in high-energy heavy-ion collisions, 
one can calculate the average polarization vector for final spin-$1/2$ hadrons with 
momentum $p$  \cite{Becattini:2013fla,Fang:2016vpj}, 
\begin{eqnarray}
P^{\mu}(p)&\equiv&\frac{d\Pi^{\mu}(p)/d^{3}p}{dN/d^{3}p} \nonumber \\
& &\hspace{-42pt}= \frac{\hbar}{4m}\frac{\int d\Sigma_{\sigma}p^{\sigma}\tilde{\Omega}^{\mu\nu}p_{\nu}\, f_{\mathrm{FD}}(x,p)[1-f_{\mathrm{FD}}(x,p)]}{\int d\Sigma_{\sigma}p^{\sigma}\, f_{\mathrm{FD}}(x,p)}, 
\label{eq:freezeout}
\end{eqnarray}
where $d\Sigma_{\sigma}$ is the differential volume vector on the Cooper-Frye hyper-surface at the hadronic  freeze-out temperature $T_{\rm f}$. We will work in the Bjorken-Milne coordinate wtih $X^\mu=(\tau,x,y,\eta)$, where $\tau=\sqrt{t^2-z^2}$ and the spatial rapidity is defined as $\tanh\eta=z/t$.

{\it (3+1)D viscous hydrodynamic model.} -- We will use the newly developed CCNU-LBNL viscous (CLVisc) hydrodynamic model with event-by-event fluctuating initial conditions from the AMPT model to calculate the $\Lambda$ spin polarization in high-energy heavy-ion collisions. The CLVisc hydrodynamic model \cite{Pang:2014ipa} is a viscous extension of an ideal (3+1)D hydrodynamic model \cite{Pang:2012he} that parallelizes the Kurganov-Tadmor (KT) algorithm \cite{Kurganov2000241} for hydrodynamic evolution and Cooper-Frye particlization on graphics processing units (GPU) using OpenCL (Open Computing Language for parallel programming of heterogeneous systems). In this study, we use time steps $\Delta \tau = 0.005$ fm, the transverse spacing $\Delta x = \Delta y = 0.1$ fm and the longitudinal spacing $\Delta \eta = 0.1$. The number of grid points are 301, 301 and 181 for the $x$, $y$ and $\eta$ direction, respectively.

The initial conditions for the CLVisc model are constructed from the AMPT model \cite{Lin:2004en} at an initial proper time $\tau_0$
 with Gaussian smearing in both transverse coordinates and spatial rapidity.  The overall normalization of the initial conditions on the energy-momentum tensor is adjusted to fit the final charged hadron multiplicity at mid-rapidity in the most central collisions and kept fixed for all other centralities \cite{Pang:2012he}.  The initial thermalization time is set at $\tau_0=0.4$ fm at the Relativistic Heavy-ion Collider (RHIC)  energies and $\tau_0=0.2$ fm at the Large Hadron Collider (LHC). The partial chemical equilibrium EoS s95p-PCE165-v0 parametrizing lattice QCD calculations \cite{Huovinen:2009yb} is used in the CLVisc model. We also assume a zero baryon chemical potential and a hadronic freeze-out temperature $T_{\rm f}=137$ MeV. In the CLVisc simulations presented in this Letter we have neglected the vorticity term in the viscous stress tensor. We also neglect viscous corrections to the final hadron spectra in the calculation of hyperon spin polarization. Both effects can be included in future studies.
 
The AMPT model uses Heavy-Ion Jet INteraction Generator (HIJING) \cite{Wang:1991hta} for the initial parton production from both incoherent semihard scatterings and coherent string formation. A string-melting mechanism is used to convert strings into partons which undergo the parton transport process before hadronization through parton coalescence.  Therefore, the AMPT initial conditions for CLVisc simulations at the initial proper time $\tau_0$ contain fluctuations in both transverse and longitudinal direction. The longitudinal variation of the initial energy density contains a systematic linear twist (rotation) and random fluctuation of the event plane, both leading to decorrelation of anisotropic flow of final hadrons with large pseudo rapidity gaps \cite{Pang:2014pxa,Pang:2015zrq}.

\begin{figure}
\centerline{\hspace{-36pt}\includegraphics[scale=0.30]{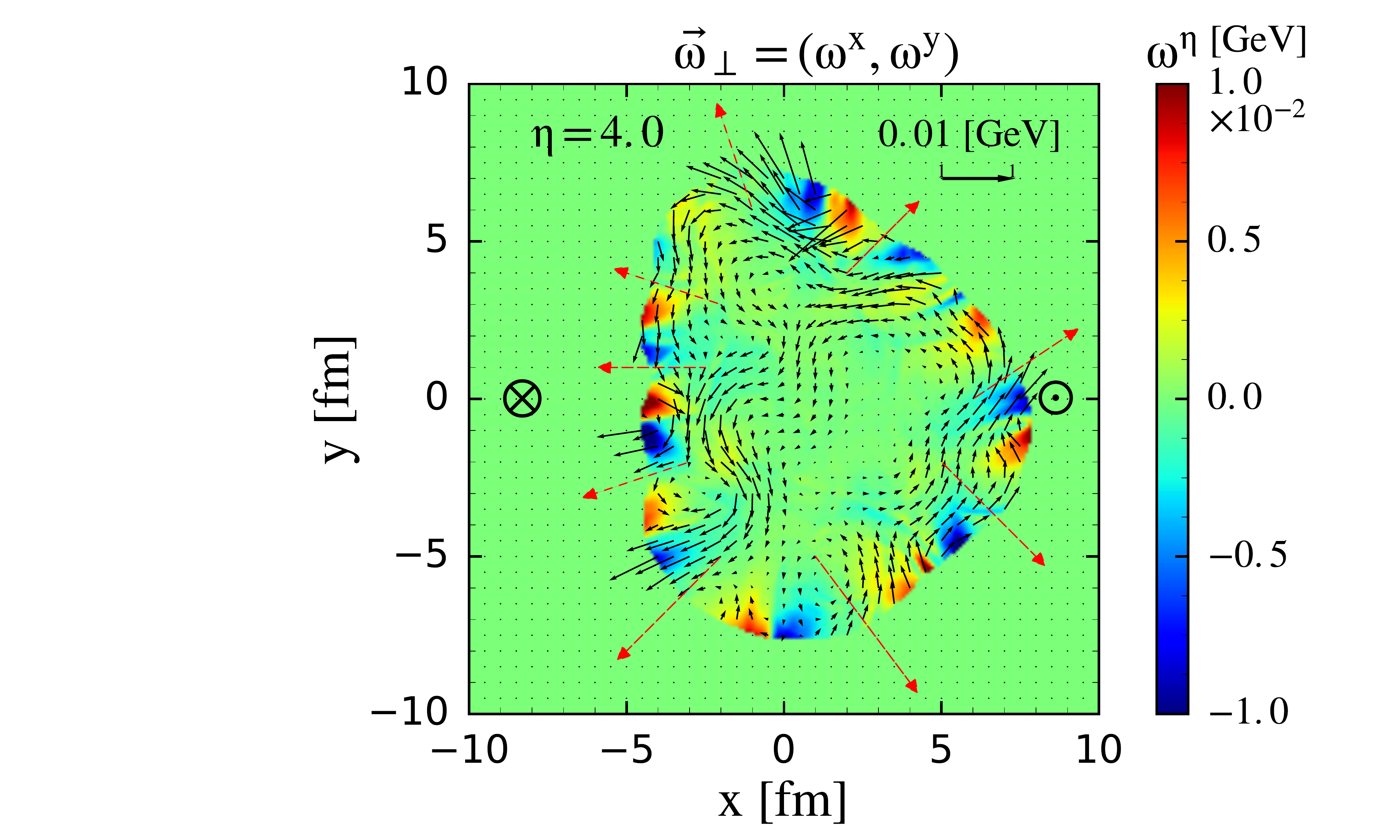}}

\vspace{-8pt}
\caption{\label{fig1}
(color online) Transverse (arrows) and longitudinal vorticity (contour) distributions in the transverse plane at $\eta=4$ in semi-peripheral (20-30\%) Au+Au collisions at $\sqrt{s}_{\rm NN}=200$ GeV with shear viscosity to entropy density ratio $\eta_v/s=0.08$. Dashed arrows indicate radial flow of hot spots.  A cut-off in energy density $\epsilon>0.03$ GeV/fm$^3$ is imposed. The direction of the beam (target) is indicated by $\odot$ ($\otimes$). The orbital angular momentum of the collision is along $-\hat y$.}
\end{figure}

{\it Convective flow and vorticity distribution.} -- The initial conditions constructed from the AMPT/HIJING model contain fluctuations in the local fluid velocity \cite{Pang:2014pxa} due to string-breaking and mini-jets. These fluctuations in fluid velocity and the energy density lead to non-vanishing local vorticity as well as global net vorticity along the orbital angular momentum of non-central collisions \cite{Deng:2016gyh}.  Collective expansion can also generate structures in the fluid velocity (radial flow and anisotropic flow for example) as well as in the vorticity.  Shown in Fig.~\ref{fig1} are distributions of the transverse $\vec\omega_\perp(x,y)$ (arrows)  and longitudinal vorticity $\omega_\eta(x,y)$ (colored contour) in the transverse plane at a spatial rapidity $\eta=4$ and proper time $\tau=3$ fm/$c$ in a semi-peripheral (20-30 \%) Au +Au collision at $\sqrt{s}_{\rm NN}=200$ GeV according to our CLVisc simulations.  One can clearly see large structures in the vorticity distributions.

According to the definition of the vorticity $\omega^\mu$, it has contributions from convection (spatial gradient of the fluid velocity), acceleration (temporal gradient of the fluid velocity) and conduction (spatial and temporal gradient of the temperature). Within CLVisc calculations, we find that the vorticity is dominated by convection. The right-hand circular structure around the beam direction (out of the transverse plane) of the transverse vorticity is essentially caused by convective longitudinal flow (transverse gradient of the longitudinal fluid velocity or longitudinal gradient of transverse fluid velocity). The total net vorticity $\langle\sum\omega_y\rangle$ projected to the reaction plane is proportional to the global orbital angular momentum and increases with the centrality of the collisions. The convective flow in the longitudinal direction is caused by both the twist and fluctuations that break boost-invariance. Therefore, the magnitude of the local transverse vorticity $\langle |\omega_\perp|\rangle$ and the net total vorticity  $\langle\sum\omega_y\rangle$ should both increase with centrality, spatial rapidity and with decreasing energy \cite{Deng:2016gyh}. Similarly, expansion of hot spots (denoted by dashed arrows in Fig.~\ref{fig1}) can also lead to convective flow in the radial direction. The right-handed vortical structure along the radial direction leads to the pairing of positive and negative longitudinal vorticity $\omega_\eta$ , or vortex-pairing, in the transverse plane at a given spatial rapidity. Since the longitudinal vorticity is caused mainly by transverse geometry and fluctuations, its magnitude and structure should depend on centrality but not on colliding energy and rapidity. The average value over the entire transverse plane $\langle\sum\omega_\eta\rangle$, however, should vanish.

\begin{figure}
\vspace{-16pt}
\centerline{\includegraphics[scale=0.25]{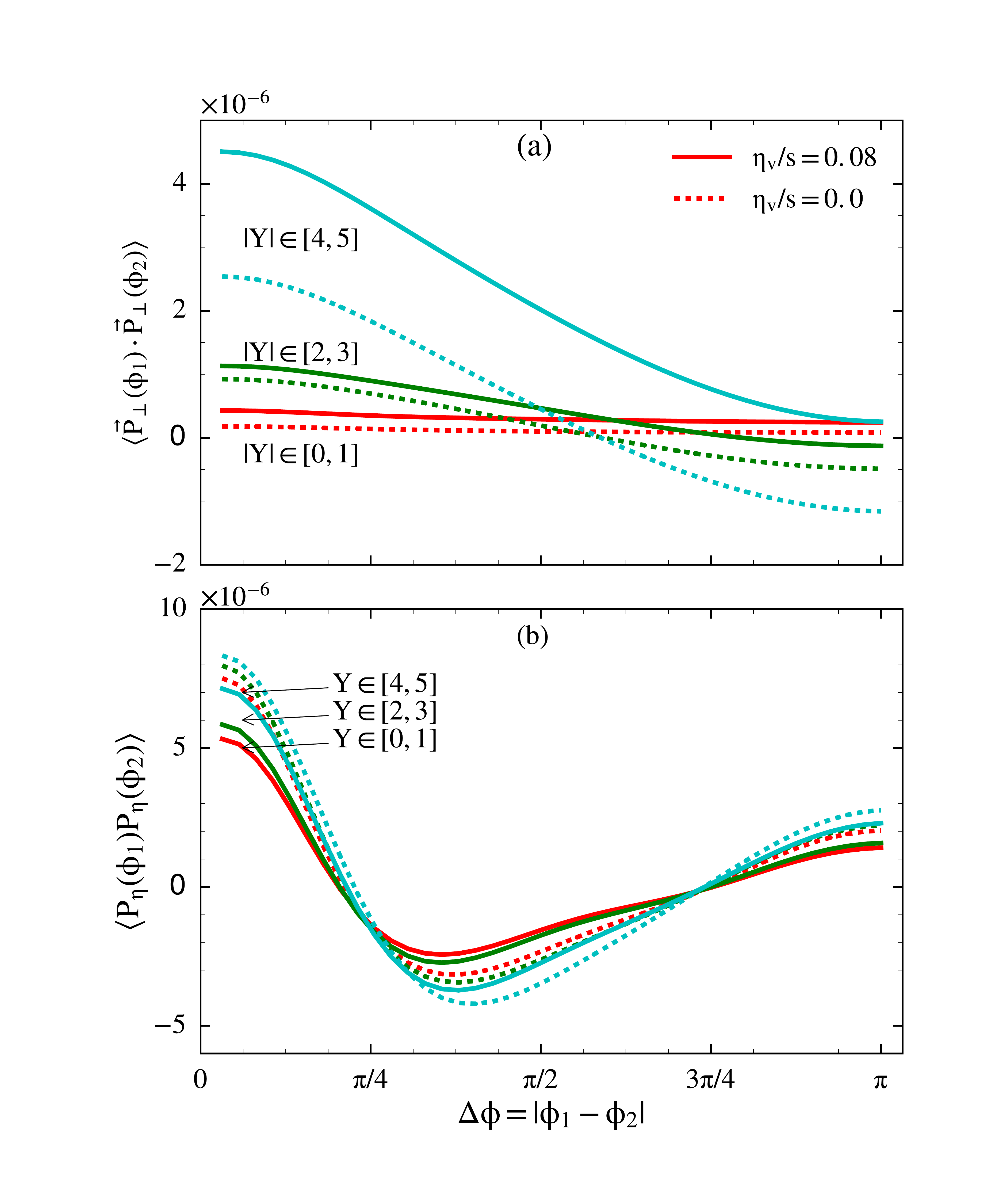}}

\vspace{-24pt}
\caption{\label{fig2}
(color online) (a) Transverse and (b) longitudinal spin correlation of two $\Lambda$'s  as a function of the azimuthal angle difference (of their momenta) in different rapidity regions of semi-peripheral (20-30\%) Pb+Pb collisions at $\sqrt{s}_{\rm NN}=2.76$ TeV with shear viscosity to entropy density ratio $\eta_v/s=0.08$ (solid) and 0.0 (dashed).}
\end{figure}

\begin{figure}
\vspace{-16pt}
\centerline{\includegraphics[scale=0.25]{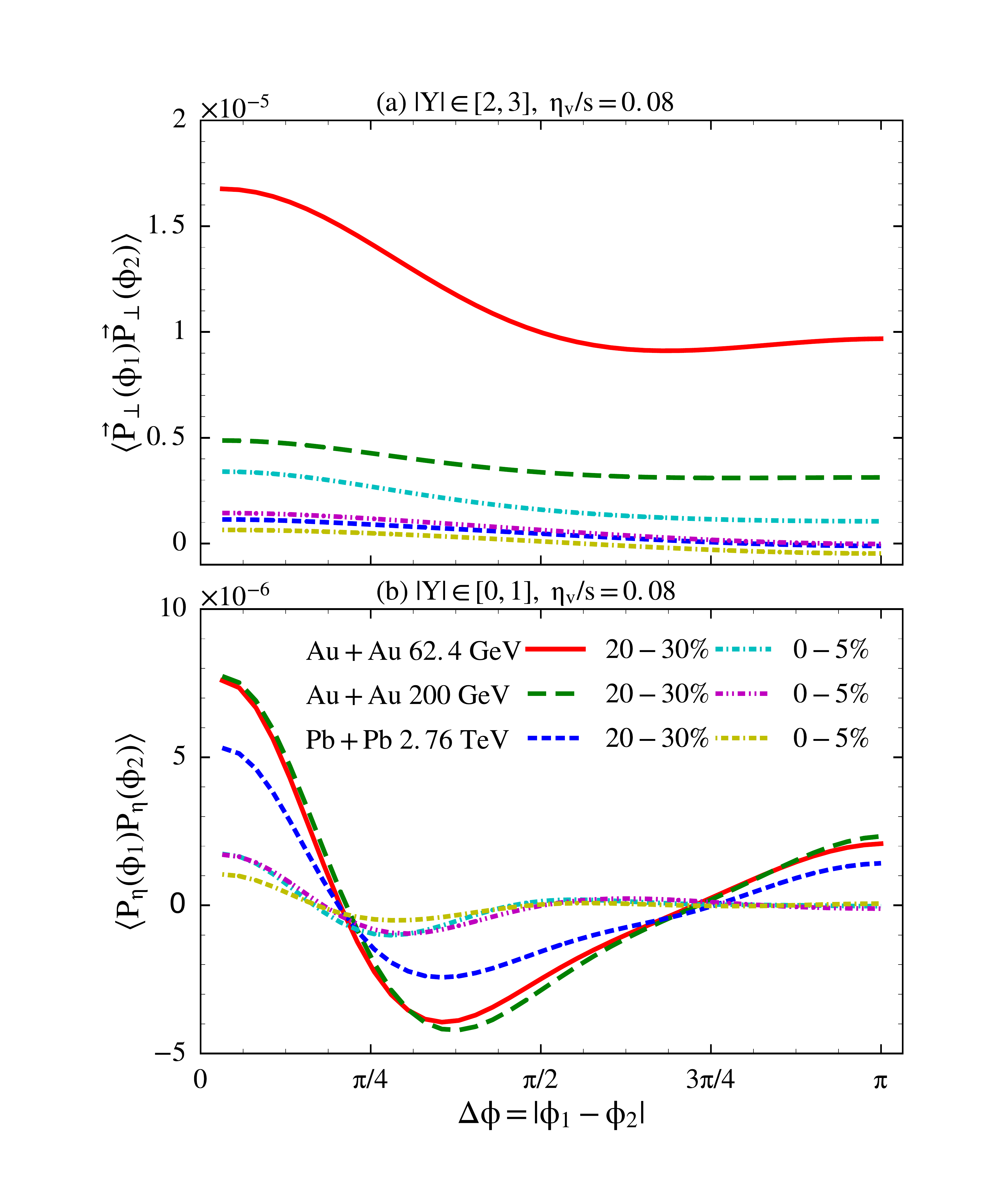}}

\vspace{-24pt}
\caption{\label{fig3}
(color online) (a) Transverse ($|Y|\in [2,3]$) and (b) longitudinal ($|Y|\in [0,1]$) spin correlation of two $\Lambda$'s  as a function of the azimuthal angle difference (of their momenta) in semi-peripheral (20-30\%) and central (0-5\%) Au+Au collisions at $\sqrt{s}_{\rm NN}=62.4, 200$ GeV and Pb+Pb collisions at $\sqrt{s}_{\rm NN}=2.76$ TeV with $\eta_v/s=0.08$.}
\end{figure}

{\it Hyperon spin correlation.} -- We propose to use the spin correlations of two $\Lambda$'s to study the vortical structure of the expanding fluid in high-energy heavy-ion collisions. Since the spin polarization is directly proportional to the local vorticity, the spatial structure in Fig.~\ref{fig1} is expected to show up in the azimuthal correlation of $\Lambda$ spin polarization due to radial expansion. Shown in Fig.~\ref{fig2} are the transverse and longitudinal spin correlations of two $\Lambda$'s , $\langle \vec P_\perp(\phi_1) \cdot \vec P_\perp(\phi_2) \rangle$ and $\langle P_\eta(\phi_1) P_\eta(\phi_2) \rangle$, respectively, as functions of the azimuthal angle difference $|\phi_1-\phi_2|$ of their momenta. In our CLVisc hydro simulations of semi-central (20-30\%)  Pb+Pb collisions at $\sqrt{s}_{\rm NN}=2.76$ TeV, we have set the shear viscosity to entropy density ratio to $\eta_v/s=0.08$ (solid lines) and 0.0 (dashed lines). As expected, the transverse spin correlation in azimuthal angle has an approximate cosine form due to the circular structure of the transverse vorticity around the beam direction plus an offset due to the global spin polarization. Both the amplitude of the oscillation (local polarization) and the offset (global polarization) increase with rapidity as well as with $\eta_v/s$.  The longitudinal spin correlation on the other hand has a different behavior. The oscillation in $|\phi_1-\phi_2|$ is the result of vortex-pairing in the transverse plane as illustrated in Fig.~\ref{fig1}.  The sign change at $|\phi_1-\phi_2|\approx 1$ indicates the typical opening angle of the vortex pairs from the convective radial flow due to transverse geometry and fluctuations.  The amplitude of the longitudinal spin correlation increases slightly with rapidity but decreases slightly with $\eta_v/s$.

In Fig.~\ref{fig3}, we show (a) the $\Lambda$ transverse spin correlations in the rapidity range $Y \in [2,3]$ and (b) the longitudinal spin correlation in $Y \in [0,1]$ in semi-peripheral (20-30\%) and central (0-5\%) Au+Au collisions at $\sqrt{s}_{\rm NN}$=62.4, 200 GeV  and Pb+Pb collisions at $\sqrt{s}_{\rm NN}$=2.76 TeV for $\eta_v/s=0.08$. Both the amplitude and the off-set of the transverse spin correlation increase with decreasing colliding energy because the transverse vorticity is bigger due to larger longitudinal fluctuations and twist at lower beam energies or larger rapidities. They also have a strong centrality dependence and become very small in central collisions.  The longitudinal spin correlation in the central rapidity region, however, does not have a strong energy dependence because of the geometric and fluctuating nature of its origin.  It has a strong dependence on the centrality. In non-central collisions, the vortex-pairing is dominated by the dipole structure of the elliptic flow and the opening angle $\Delta\phi\sim 3\pi/8 $ indicates the scale of the vortex pair. In central collisions, the vortex-paring is dominated by hot spots and the angular structure of the longitudinal spin correlations indicates the size and distance between these hot spots. 

{\it Summary and discussions.} -- We have studied the vortical structure of the sQGP fluid in high-energy heavy-ion collisions using the CLVisc hydrodynamic model with fluctuating initial conditions from the AMPT/HIJING model. The transverse vorticity has a circular structure around the beam direction in addition to the average net vorticity along the reaction plane due to global orbital angular momentum in non-central collisions. The longitudinal vorticity has a vortex-pairing structure in the transverse plane. We propose to use the $\Lambda$ spin correlations to measure these vortical structures. We predict both the transverse and longitudinal spin-correlation as functions of the azimuthal angle difference of  two $\Lambda$s' momenta. Measurements of these spin correlation functions can give us a detailed picture of the flow and vortical structure and provide important constraints on the initial condition and the transport properties of the dense matter in high-energy heavy-ion collisions. We limit our predictions to high-energy heavy-ion collisions where one can neglect the finite baryon chemical potential. For collisions at the beam scan energies at RHIC, one has to include baryon number conservation and an EoS with finite baryon chemical potential in hydrodynamic simulations. We also did not include the vorticity term in the shear stress tensor and viscous corrections to the phase-space distribution of final hadrons. We have neglected the effect of the magnetic field on the spin polarization which should be small during the hadronic freeze-out following equilibrium hydrodynamic evolution. These effects can be addressed in more accurate studies in the future.

\textit{Acknowledgments.} -- This work is supported in part 
by NSFC under the Grant Nos. 11221504 and 11535012, by MOST of China under Grant No. 2014DFG02050, 
by the Major State Basic Research Development Program (MSBRD) in China under the Grant No. 2015CB856902,
2014CB845404, and 2014CB845406, by U.S. DOE under Contract No. DE-AC02-05CH11231, by Helmholtz Young Investigator Group VH-NG-822 from the Helmholtz Association and GSI and by HIC for FAIR within the framework of the Landes-Offensive zur Entwicklung Wissenschaftlich-Oekonomischer Exzellenz (LOEWE) program launched by the State of Hesse. Computations are performed at the Green Cube at GSI and GPU workstations at CCNU.

\bibliographystyle{apsrev}
\addcontentsline{toc}{section}{\refname}\bibliography{ref-spin}

\end{document}